\documentclass[sigconf,natbib]{acmart}

\AtBeginDocument{%
  }

\copyrightyear{2025}
\acmYear{2025}
\setcopyright{acmlicensed}\acmConference[SIGIR '25]{Proceedings of the 48th International ACM SIGIR Conference on Research and Development in Information Retrieval}{July 13--18, 2025}{Padua, Italy}
\acmBooktitle{Proceedings of the 48th International ACM SIGIR Conference on Research and Development in Information Retrieval (SIGIR '25), July 13--18, 2025, Padua, Italy}
\acmDOI{10.1145/3726302.3730235}
\acmISBN{979-8-4007-1592-1/2025/07}
\usepackage{multirow}
\usepackage{caption}
\usepackage{subcaption}
\usepackage{graphicx}
\usepackage{float} 
\begin{document}

\title{NAM: A Normalization Attention Model for Personalized Product Search In Fliggy}


% \author{
% Shui Liu$^{1}$,
% Mingyuan Tao$^{1}$, 
% Maofei Que$^{1}$, 
% Pan Li$^{2}$, 
% Dong Li$^{1}$,
% Shenghua Ni$^{1}$,
% Zhuoran Zhuang$^{1}$\\
% % 机构列表
% $^{1}$ Alibaba Group,
% $^{2}$ Georgia Tech\\ 
% \{shui.lius, mingyuan.tmy, maofei.qmf, nantian\}@alibaba-inc.com; \\
% \{shiping, shenghua.nish\}@taobao.com;  pan.li@scheller.gatech.edu
% }

% ORCIDs
%
% Shui Liu : 0000-0001-5189-1841
% Mingyuan Tao : 0000-0001-7710-3401
% Maofei Que: 0009-0004-1410-3825
% Pan Li: 0000-0003-4957-3064
% Dong Li: 0000-0002-4715-9479
% Shenghua Ni：0000-0001-7172-6077
% Zhuoran Zhuang: 0009-0003-7028-6067

% \author{Shui Liu}
% \affiliation{%
%   \institution{Alibaba Group}
% }
% \email{shui.lius@alibaba-inc.com}

% \author{Mingyuan Tao}
% \affiliation{%
%   \institution{Alibaba Group}
% }
% \email{mingyuan.tmy@alibaba-inc.com}

% \author{Maofei Que}
% \affiliation{%
%   \institution{Alibaba Group}
% }
% \email{maofei.qmf@alibaba-inc.com}

% \author{Pan Li}
% \affiliation{%
%   \institution{Georgia Tech}
% }
% \email{pan.li@scheller.gatech.edu}

% \author{Dong Li}
% \affiliation{%
%   \institution{Alibaba Group}
% }
% \email{shiping@taobao.com}

% \author{Shenghua Ni}
% \affiliation{%
%   \institution{Alibaba Group}
% }
% \email{shenghua.nish@taobao.com}

% \author{Zhuoran Zhuang}
% \affiliation{%
%   \institution{Alibaba Group}
% }
% \email{nantian@alibaba-inc.com}

\author{Shui Liu}
\email{shui.lius@alibaba-inc.com}
% \affiliation{%
%   \institution{Alibaba Group}
%   % \city{Hangzhou}
%   % \state{Zhejiang}
%   % \country{China}
% }
\author{Mingyuan Tao}
\email{mingyuan.tmy@alibaba-inc.com}
% \affiliation{%
%   \institution{Alibaba Group}
%   % \city{Hangzhou}
%   % \state{Zhejiang}
%   % \country{China}
% }
\author{Maofei Que}
\email{maofei.qmf@alibaba-inc.com}
\affiliation{%
  \institution{Alibaba Group}
  \city{Hangzhou}
  \state{Zhejiang}
  \country{China}
}

\author{Pan Li}
\email{pan.li@scheller.gatech.edu}
\affiliation{%
  \institution{Georgia Tech}
  \city{Atlanta}
  \state{Georgia}
  \country{USA}
}

\author{Dong Li}
\email{shiping@taobao.com}
% \affiliation{%
%   \institution{Alibaba Group}
%   % \city{Hangzhou}
%   % \state{Zhejiang}
%   % \country{China}
% }

\author{Shenghua Ni}
\email{shenghua.nish@taobao.com}
% \affiliation{%
%   \institution{Alibaba Group}
%   % \city{Hangzhou}
%   % \state{Zhejiang}
%   % \country{China}
% }

\author{Zhuoran Zhuang}
\email{nantian@alibaba-inc.com}
\affiliation{%
  \institution{Alibaba Group}
  \city{Hangzhou}
  \state{Zhejiang}
  \country{China}
}

% \renewcommand{\shortauthors}{Liu, et al.}
\renewcommand{\shortauthors}{Shui Liu et al.}
%% No italics, no superscripts
%% Use footnote or author note to identify equal contribution and/or contact author info

% \renewcommand{\abstractname}{ABSTRACT}
% \renewcommand{\keywordsname}{KEYWORDS}

\begin{abstract}

Personalized product search provides significant benefits to e-commerce platforms by extracting more accurate user preferences from historical behaviors. Previous studies largely focused on the user factors when personalizing the search query, while ignoring the item perspective, which leads to the following two challenges that we summarize in this paper: First, previous approaches relying only on co-occurrence frequency tend to overestimate the conversion rates for popular items and underestimate those for long-tail items, resulting in inaccurate item similarities; Second, user purchasing propensity is highly heterogeneous according to the popularity of the target item: it is less correlated with the user's historical behavior for a popular item and more correlated for a long-tail item.
To address these challenges, in this paper we propose \textbf{NAM}, a \textbf{N}ormalization \textbf{A}ttention \textbf{M}odel, which optimizes ''when to personalize'' by utilizing Inverse Item Frequency (IIF) and employing a gating mechanism, as well as optimizes ''how to personalize'' by normalizing the attention mechanism from a global perspective. Through comprehensive experiments, we demonstrate that our proposed NAM model significantly outperforms state-of-the-art baseline models. Furthermore, we conducted an online A/B test at Fliggy, and obtained a significant improvement of 0.8\% over the latest production system in conversion rate. 

\end{abstract}






\maketitle

\section{Introduction}

% product search personalization benifits a lot , but as xxx mentioned , whether to p13n should be studied.

% when and how to p13n , is to study when and how to use user bhv history 

% previous work , through Analysis of query， 

% Product search is one of the most important components of e-commerce websites, fulfilling a significant portion of users' needs for product discovery.
Personalized product search addresses the critical problem of information overload in industrial practices \cite{querytaxonomy,intentsatisfaction}. It operates by allowing different individuals to receive personalized results for the same query according to their historical behaviors. This significantly enhances the relevance of search results to user preferences and greatly improves product retrieval performance. Despite its great potential, personalization might not always be beneficial \cite{p13n_or_not,PotentialandPitfalls,short_long_term_p13n}, since customers’ purchase decisions are affected by both the relevance of target items and the customer preferences. Therefore, it is crucial to understand both ''when to personalize'' and ''how to personalize'' when developing the search model. To that end, a series of deep-learning models have been proposed to extract user preferences from behavior sequences based on the current search context, such as \cite{din,dien,zeroattention,ALSTP}, where attention mechanism plays a crucial role in all these models. Specifically, attention variant models that focus on the relationship between queries and user behavior sequences have been proposed\cite{zeroattention,denoiseattention,tem}, allowing the weight of personalization to be as large as 1 when personalization is dominant and as small as 0 when it is not needed at all.

% \textcolor{red}{To address this issue, ZAM introduces a zero vector, allowing attention weights to be assigned to this zero vector in scenarios where personalization is not necessary, effectively ignoring user-specific preferences. DAM\cite{denoiseattention} suggests that negative attention weights of items in user historical behaviors to the query become positive weights after applying softmax, introduce noises. It proposes using cosine similarity instead of scaled dot product, along with ReLU\cite{relu}, while adding a similarity threshold hyper-parameter to the attention weight, to remove the influence of the irrelevant items. TEM holds that in ZAM, the importance of personalization is at most equivalent to that of the query. But in certain cases, personalization could have a greater impact than the queries themselves. Therefore, it incorporates the query vector into the user behavior sequence, allowing the importance of personalization as large as 1 and as small as 0.}

However, the research gap lies in the fact that, to the best of our knowledge, previous models are all built upon query-dependent attention and tend to ignore the characteristics of items as a result. We argue that exploring personalization from the item perspective is particularly important due to the following reasons: (1) To answer the question of ''when to personalize'', the popularity of the items needs to be taken into account, since customers’ purchase decision is also affected by the bandwagon effect \cite{unfairness_of_popularity,matthew_effect} brought by the item's popularity. (2) Relying only on co-occurrence item frequency will lead to the overestimation of conversion rates for popular items and the underestimation for long-tail items, resulting in a Matthew effect\cite{how_popular_bias}, as well as the suboptimal item similarities or attention weights.
Considering these challenges, we propose three intuitions, which motivate us to conduct preliminary data analysis to validate them on our industry dataset. Based on the data analysis, we propose the Normalization Attention Model (NAM), which consists of a gating mechanism with IIF of target item to model when to personalize, as well as a global normalized attention mechanism with IIF of items in user historical behaviors to model how to personalize.

% Then we demonstrate NAM outperforms existing models through offline evaluations and observe improvement on conversion rate by online A/B test.

% \input{sections/relatedwork}

\section{Intuition \& Data Analysis}

In this section, we present three intuitions and demonstrate them through data analysis, which inspired the methods for enhancing the ranking model.

\textbf{Intuition 1: Popular items are overestimated, while long-tail items are underestimated.}
Due to the fact that the search model is trained based on historical data, popular items dominate the model, resulting in the model overestimating the click-through rate for popular items while underestimating the click-through rate for long-tail items. To demonstrate this intuition, we collect consumer logs from the product search system on Fliggy\footnote{\url{http://www.fliggy.com/}}. Based on the logs, we sort items in descending order based on the number of users who interacted with them and divide them into five levels, then compare the predicted click-though rate \& post-click conversion rate (pCTCVR), the actual click-though rate \& post-click conversion rate (CTCVR), and predicted Conversion rate Over Conversion rate (PCOC) corresponding to each level. PCOC is calculated as: $PCOC = \frac{pCTCVR}{CTCVR}$. The result is presented in Table \ref{tab:pcoc_baseline}, which clearly indicates that popular items are overestimated, while long-tail items are underestimated.

% TODO: how we conduct the data analysis 

\begin{table}[htbp]
  \caption{PCOC on Different Item Popularity Level. The closer the value of PCOC is to 1, the better.}
  \resizebox{0.47\textwidth}{!}{
  \label{tab:pcoc_baseline}
  \begin{tabular}{c c c c c }
    
    \hline\hline
    {Level} & {ExposurePercentage} & {pCTCVR} & {CTCVR} & {PCOC} \\
    \hline

    {[0,1\%]     }&{	57.08\%}&	{0.74\%}&	{0.63\%}&	{1.1715} \\ 
    {(1\%,5\%]	  }&{   21.29\%}&	{0.79\%}&	{0.79\%}&	{0.9957} \\ 
    {(5\%,20\%]  }&{	16.26\%}&	{0.71\%}&	{0.79\%}&	{0.9076} \\ 
    {(20\%,100\%]}&{	5.37\% }&   {0.18\%}&	{0.22\%}&	{0.8174} \\ 
    \hline
    \end{tabular}
    }
\end{table}

\textbf{Intuition 2: Globally-normalized item similarity leads to better recommendation results.}
\label{intuition2}
Proper suppression of popular items improves the performance of the recommender system. For example, adjusting the weight of item frequency in the similarity formula is often used to alleviate the Harry Potter problem\cite{harry_potter} in recommendation system\cite{iuf}. Based on consumer logs collected from our product search system, we investigate the hit rate with different similarity formulas. The hit rate metric is calculated as:
% \begin{equation}
%     HitRate@k = \frac{\sum\limits_{u \in U}{\sum\limits_{i \in I_u}{\#\{\Gamma_i \cap S_{i,k}\}}}} {\sum\limits_{u \in U} \sum\limits_{i \in I_u}{\#\Gamma_i}}
% \end{equation}
$
    HitRate@k = \frac{\sum\limits_{u \in U}{\sum\limits_{i \in I_u}{\#\{\Gamma_i \cap S_{i,k}\}}}} {\sum\limits_{u \in U} \sum\limits_{i \in I_u}{\#\Gamma_i}}
$, where u represents a user, i represents an item. $I_u$ is the item set that interacted with u. $\Gamma_i$ is the collection of items that have interactions occurring later than item i. $S_{i,k}$ is the collection of top k similar items of item i. "$\#$" represents the number of elements in a set. Co-occurrence similarity is calculated as: $Co\text{-}occurrenceSim_{\ i,j} = \#(U_i \cap U_j ) $ , where $U_i$ is the users that interacted with item $i$. Jaccard similarity is calculated as: $JaccardSim_{i,j} = \frac{\#(U_i \cap U_{j})}{\#(U_i \cup U_j)}$ , and Cosine similarity is calculated as: $CosineSim_{i,j} =  \frac{\#(U_i \cap U_{j})}{\sqrt{\#U_i \#U_j}}$.
It's evident from Table \ref{tab:hitrate} that the hit rate of Jaccard and Cosine similarity performs much better than Co-occurrence Frequency similarity, since Jaccard and Cosine similarity involves the item's frequency as global normalization.

\begin{table}[htbp]
  \caption{Hit Rate with different Similarity Methods.}
  \resizebox{0.47\textwidth}{!}{
  \label{tab:hitrate}
  \begin{tabular}{c | c c c }
    
    \hline\hline
    Method & {HitRate@10} & {HitRate@20} & {HitRate@30} \\
    \hline
    {Co-occurrence Frequency }&{	10.04\%}&	{17.27\%}&	{22.61\%} \\ 
    {Jaccard	  }&{ 11.75\%}&	{20.90\%}&	{28.24\%} \\ 
    {Cosine }&{	12.23\%}&	{22.3\%}&	{30.33\%}\\ 
    \hline
    \end{tabular}
    }
\end{table}

\textbf{Intuition 3: Compared to popular items, the conversion rate of long-tail items is more correlated with the user's historical behaviors.}
The data in Table \ref{tab:retarget_and_conversion} shows that for long-tail items with a popularity level in the range of $(20\%,100\%]$, the conversion rate of items that users interacted with 
(IsRetarget=1) is 19.72 times higher than that of items they did not interact with (IsRetarget=0). In contrast, for the top 1\% of popular items, the conversion rate of items that users interacted with is 11.5 times higher than that of items they did not interact with, which is significantly lower than the ratio observed for long-tail items. Therefore, we can conclude that whether an item converts is more heavily influenced by similar items in user historical behaviors when the item is a long-tail item. In other words, if the target item is a long-tail item, user historical behavior should be paid more attention.

\begin{table}
  \caption{Comparison of conversion rates between retargeted and non-retargeted items on different levels of item popularity.}
  \resizebox{0.47\textwidth}{!}{
  \label{tab:retarget_and_conversion}
  \begin{tabular}{c c c c c }
    
    \hline\hline
    {Level} & {IsRetarget} & {ExposureRate} & {CTCVR} & {Retarget/NonRetarget} \\
    \hline
    \multirow{2}*{(0,1\%]} & 0 & 33.95\% & 0.50\% & \multirow{2}*{10.50} \\ 
    & 1 & 22.45\% & 5.20\% &\\ 
    
    \hline
    \multirow{2}*{(1\%,5\%]}	    & 0	& 12.86\%  	&0.58\% & \multirow{2}*{11.58}\\	
    	    & 1	& 9.57\%	&6.77\% & 	 \\
    \hline
    \multirow{2}*{(5\%,20\%]}    & 0	& 9.51\%	&0.57\% & \multirow{2}*{11.27}\\	
                & 1	& 7.86\%	&6.46\% & 	 \\
    \hline
    \multirow{2}*{(20\%,100\%]}  & 0	& 3.15\%	&0.17\% & \multirow{2}*{19.72}\\	
                & 1	& 0.65\%	&3.36\% & 	 \\
    
    \hline
    \end{tabular}
    }
\end{table}

\section{Proposed Model}
% In this section, we introduce our proposed NAM model, as presented in Fig \ref{fig:model}., while we discuss the details of the Global Normalized Multi-Head Self-Attention module (GN\_MHSA) in Section \ref{gn_mhsa}. Global Normalized Target Attention (GN\_TA) in Section \ref{gn_ta}, and IIF-based personalization in Section \ref{iif_p13n_section} respectively.

\begin{figure*}
  \centering
  \includegraphics[width=0.5\linewidth]{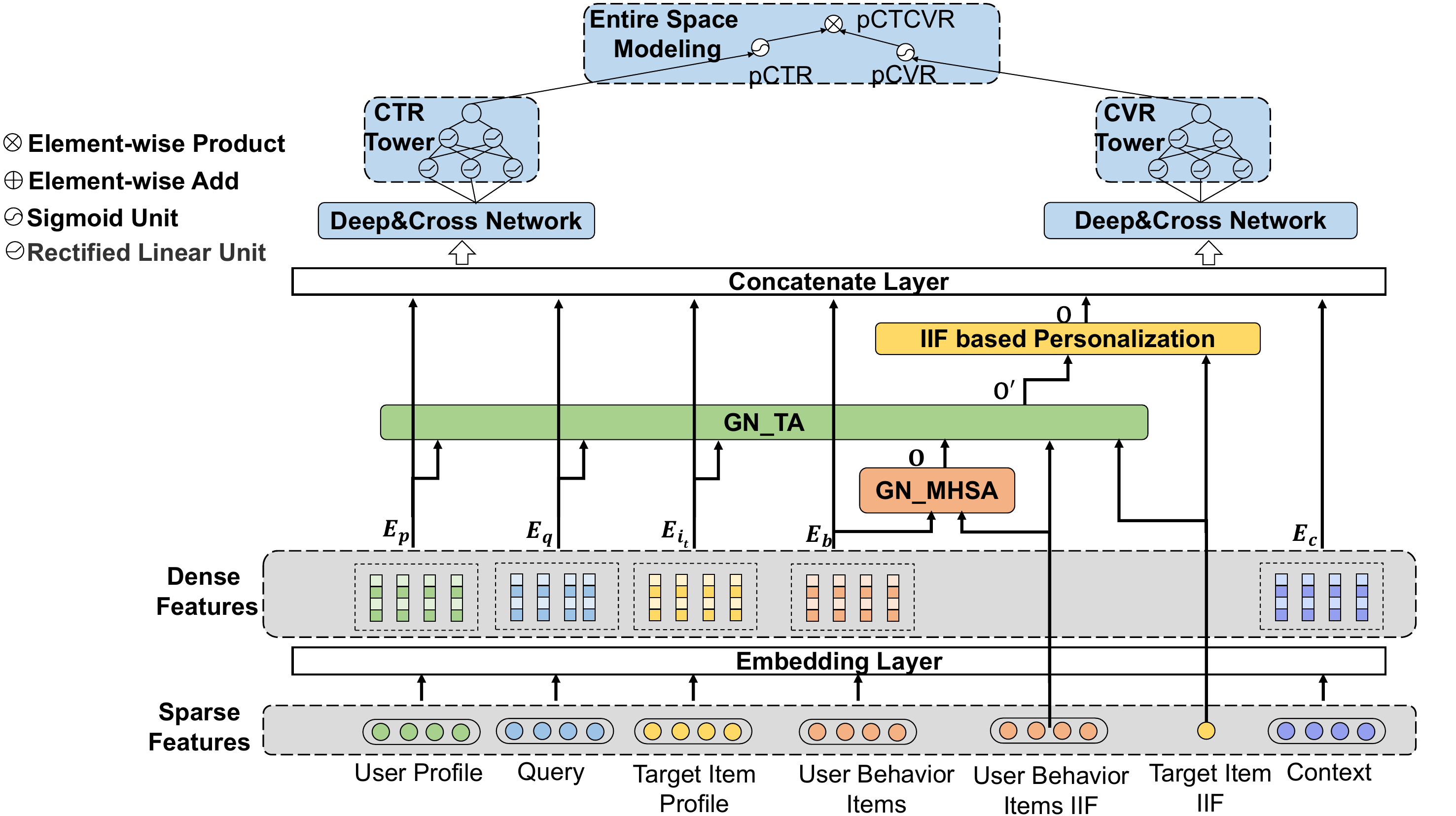}
  \caption{The overview of NAM, which consists of Embedding Layer, Global Normalized Multi-Head Self-Attention module (GN\_MHSA), Global Normalized Target Attention module (GN\_TA), IIF-based Personalization module, Concatenate layer , CTR tower \& CVR tower and Entire Space Modeling.}
  \label{fig:model}
\end{figure*}

\subsection{Global Normalized Multi-Head Self-Attention(GN\_MHSA) }
\label{gn_mhsa}
To alleviate the Matthew Effect caused by popular items being overestimated and long-tail items being underestimated, we propose the global normalized mechanism, which incorporates the square root of the product of the IIF of item $i$ and item $j$ when calculating the attention weight between item $i$ and item $j$, as Illustrated in Eq.(\ref{globalnorm}). Such square root is the denominator part of the cosine similarity formula, as mentioned in Intuition \ref{intuition2}. Each head of GN\_MHSA is calculated as follows:
% \begin{equation}
% e_{i} = EmbeddingLayer(x_{i}) \\ 
% \end{equation}
\begin{equation}
E_{b} = Concat([e_{i_1},....,e_{i_n}]) \\ 
\end{equation}
\begin{equation}
Q_h = E_bW_h^Q \ ,\ K_h = E_bW_h^K \ ,\ V_h = E_bW_h^V \\ 
\end{equation}
\begin{equation}
IIF_Q = IIF_K = [[\frac{1}{\#U_{i_1}}],......, [\frac{1}{\#U_{i_n}}]] \\
\end{equation}
\begin{equation}
\begin{split}
M_{Q,K} = SiLU(\sqrt{IIF_Q (IIF_K)^T} \odot W^{M_{Q,K}})\\  
\end{split}
\end{equation}
\begin{equation}
\label{globalnorm}
GlobalNorm_h^{MHSA} = \frac{Q_h{K_h}^T}{\sqrt{d_k}} \odot M_{Q,K}\\
\end{equation}
\begin{equation}
\begin{split}
Head_h = Softmax(GlobalNorm_h^{MHSA})V_h \\ 
\end{split}
\end{equation}
\begin{equation}
O = Concat([Head_{1},...,Head_{h},...,Head_{H}])
\end{equation}
where $i$ is one of the items in the user's historical behaviors, $n$ is the length of the user's behavior sequence, and $e_i$ is the feature embedding of item $i$. 
$E_b \in \mathbb{R}^{n \times d_{e_i}}$ is the embedding of the  user behavior sequence. 
$W_h^Q \in \mathbb{R}^{d_{e_i} \times d_k}, W_h^K \in \mathbb{R}^{d_{e_i} \times d_k}, W_h^V \in \mathbb{R}^{d_{e_i} \times d_v}, W^{M_{Q,K}} \in \mathbb{R}^{1 \times 1} $ are learnable model parameter matrices, $d$ indicates dimension. $IIF_{i}$ of item $i$ is calculated as :$IIF_i = \frac{1}{\#U_{i}}$ , where $U_i$ is the number of users that interacted with item $i$. $SiLU$ \cite{silu} is the swish activation function, we use $SiLU$ instead of $Sigmoid$ or $ReLU$ here to ensure that the values of $M_{Q,K}$ are within the range of 0 to 1 since the values of IIF are always positive. We denote $O$ as the output of GN\_MHSA module, where $H$ is the number of heads.

\subsection{Global Normalized Target Attention(GN\_TA) }
We also applied global normalization to the target attention module. The target attention in GN\_TA module takes embedding of context features as Q part of the attention mechanism, and takes $O$ (the output of GN\_MHSA) as K/V part of the attention mechanism. Context features consist of user features, target item features, and query features. $O'$, denoted as the output of $GN\_TA$, is calculated as follows: 

\label{gn_ta}
% \begin{equation}
%     E_{i_t} = EmbeddingLayer(x_{i_t}) \\  
% \end{equation}

\begin{equation}
    C = Concat([{E}_{user},{E}_{query},{E_{i_t}}]) \\
\end{equation}
\begin{equation}
    IIF_{i_t} = [[\frac{1}{\#U_{i_t}}]] \\ 
\end{equation}
\begin{equation}
    M_{i_t,K'} = SiLU(\sqrt{IIF_{i_t} (IIF_K)^T}\odot W^{M_{i_t,K'}}) \\ 
\end{equation}
\begin{equation}
    Q' = CW^{Q'} \ ,\ K' = OW^{K'}\ ,\ V' = OW^{V'} \\ 
\end{equation}
% \begin{equation}
%     K' = UW^{K'} \\ 
% \end{equation}
% \begin{equation}
%     V' = UW^{V'} \\ 
% \end{equation}
\begin{equation}
    GlobalNorm^{TA} = \frac{Q'{K'}^T}{\sqrt{d_{K'}}} \odot M_{i_t,K'}\\ 
\end{equation}
\begin{equation}
    O' = Pooling(Softmax(GlobalNorm^{TA})V') \\ 
\end{equation}
where $E_{user}$ is embedding of user features , $E_{query}$ is embedding of query features, and $E_{i_t}$ is embeddings of target item features. $C$ represents the embedding of context features. $W^{M_{i_{t},K'}} \in \mathbb{R}^{1 \times 1}, W^{Q'} \in \mathbb{R}^{d_C \times d_{k'}}, W^{K'} \in \mathbb{R}^{d_{O} \times d_{k'}}, W^{V'} \in \mathbb{R}^{d_{O} \times d_{v'}} $ are learnable model parameter matrices.
% \begin{equation}
%     U' = LayerNorm(TA\_output) 
% \end{equation}

\subsection{IIF-based Personalization}
\label{iif_p13n_section}
Regarding the question of when to personalize, we propose a gating layer based on IIF to automatically balance the influence of personalization: when an item is very popular, we reduce the dependence on users' historical preferences, thereby decreasing the influence of personalization. Conversely, for long-tail items, the influence of personalization will be increased. The gating layer formula is described as follows: 
\begin{equation}
\label{iif_p13n}
    \begin{aligned}
        O'' = SiLU(IIF_{i_t}W^{IIF_{i_t}})  \odot O'
    \end{aligned}
\end{equation}
where $W^{IIF_{i_t}} \in \mathbb{R}^{1 \times 1}$ is a learnable model parameter.

% \subsection{Model Optimization}

\section{Experiments}

\subsection{Dataset and Experimental Settings}
\textbf{Dataset Descriptions.} We collect consumer logs from the product search system in Fliggy as the offline dataset. The statistics of the dataset are presented in Table \ref{tab:dataset_desc}. The dataset contains about 374 million training samples and 9 million test samples, which are divided by date.

\begin{table}[htbp]
    \centering
    \caption{Statistics of the Offline Dataset}
    \begin{tabular}{ccccc}
        \hline \hline
        \# User & \# Item & \# Impression & \# Click & \# Conversion\\ 
        \hline
          6,267,425 
          & 975,170 
          & 383,224,804
          & 34,106,927
          & 2,766,688 \\
         \hline
    \end{tabular}
    \label{tab:dataset_desc}
\end{table}

\textbf{Baseline Models}: 
(1) \textbf{QEM} \cite{hem}, which computes query embeddings by encoding query words with a nonlinear projection function.
(2) \textbf{HEM} \cite{hem}, which takes into account both the query and user preferences, where user preferences are represented by the text embeddings of items that the user has interacted with. In this paper, we use the clicked item’s titles as the text associated with each user.
(3) \textbf{AEM} \cite{zeroattention}, which models the relationship between the query and user behavior sequence with the classical attention mechanism.  The NAM is equivalent to the AEM when the Global Normalized Attention module and IIF-based personalization module are removed.
(4) \textbf{ZAM} \cite{zeroattention}, which introduces a zero vector to AEM to conduct differentiated personalization for product search based on both the query characteristics and user purchase histories.
(5) \textbf{TEM} \cite{tem}, which encodes the sequence of the query and users’ purchase history with a transformer architecture, allowing the effect of personalization to vary from none to domination.
(6) \textbf{DAM} \cite{denoiseattention}, which uses cosine similarity instead of scaled-dot product, along with ReLU\cite{relu}, while adding a similarity threshold hyper-parameter to the attention weight, to remove the influence of the irrelevant items. 

\textbf{Experimental Settings.} To ensure a fair comparison, our proposed model and all baseline models share the same network structure and hyperparameters, where we i) share the same features; ii) use a three-layer fully connected network with sizes of 512, 256, and 128 to implement the CTR/CVR towers, and apply batch normalization to each layer for convergence; iii) employ $H=4$ as the number of self-attention heads and $d_{e_i}=98$ as the dimension of item embedding; iv) adopt ADAM optimizer to train all models, with a fixed learning rate of 0.001 and a batch size of 1024.
% TODO : num of heads and d_{e_i} and d_k,d_v

\textbf{Metrics.} We employ two widely used metrics to measure the offline performance of the models\cite{din,dien,esmm,mmoe,PLE,esm2}: Area Under Curve(AUC)\cite{auc} and Group Area Under Curve(GAUC) \cite{gauc}.
\subsection{Offline Comparison Results}
Table \ref{table:auc} shows offline evaluation results of different models on our industrial dataset, where our proposed NAM model significantly outperforms all baselines on both CTCVR AUC and CTCVR GAUC. Compared to the second-best baseline, NAM achieved a 0.001 improvement on CTCVR AUC and 0.002 improvement in CTCVR GAUC. In comparison to AEM, the baseline model in our online environment, NAM achieved a 0.0027 improvement in CTCVR AUC and 0.004 improvement in CTCVR GAUC. To validate the effectiveness of different components of NAM, we remove global normalization and IIF based personalization of NAM and make a comparison with our NAM model. As shown in Table \ref{table:auc}, global normalization and IIF-based personalization are both useful for performance improvements. 

\begin{table}[htbp]
    \centering
    \caption{Comparison of models on the offline industry dataset. Results of CTR, and CTCVR are presented. The best results of all methods are indicated in bold and second-best results are indicated in underline. The Improvement means the AUC/GAUC improvement of NAM compared with the second-best. P13n is the abbreviation for personalization.}
    \resizebox{0.4\textwidth}{!}{
    \begin{tabular}{ccccc}
    \hline\hline
    \multirow{2}*{Models}&\multicolumn{2}{c}{AUC}&\multicolumn{2}{c}{GAUC}\\
    \cmidrule(r){2-3} \cmidrule(r){4-5}
    &CTR& CTCVR&CTR &CTCVR  \\ 
    \hline
    QEM     &0.7621	&0.9032 &0.7126		&0.8181 \\
    
    HEM     & 0.7626	&0.9034       &0.7121		&0.8179	  \\
    AEM (Base)     & 0.7615    &0.9041       &0.7148       &0.8205   \\
    ZAM     & 0.7647	&0.9055       &\underline{0.7175}		&\underline{0.8225}   \\
    TEM     & \underline{0.7651}         &\underline{0.9058}	          &0.7166           &0.8217	      \\
    DAM     & 0.7644	&0.9049       &0.7158		&0.8201	  \\
    \hline
    NAM w/o IIF based P13n &0.7633	    &0.9054 &0.7184		&0.8232\\
    NAM w/o GlobalNorm & 0.7621	&0.9057 &0.7177		&0.8218\\
    NAM     & \textbf{0.7665}	& \textbf{0.9068} & \textbf{0.7192}	& \textbf{0.8245}\\
    \hline 
    Improvement     & +0.0014 	&+0.001  &+0.0017 	& +0.002\\
    \hline\hline
    \end{tabular}
    }
\label{table:auc}

\end{table}
\subsection{PCOC Analysis}
Table \ref{tab:nam_pcoc} shows the comparison of PCOC between NAM and AEM on different item popularity level. Compared to AEM, the PCOC of NAM decreases by 3.87\% for popular items in the popularity level range of (0, 1\%] and increases by 2.69\% for long-tail items in the popularity level range of (20\%, 100\%]. This indicates that our model has made some progress in alleviating the issue of popular items being overestimated and long-tail items being underestimated.

\begin{table}[H]
  \caption{Comparison of PCOC between NAM and AEM on Different Item Popularity Level.  The closer
the value of PCOC is to 1, the better.}
  \resizebox{0.25\textwidth}{!}{
  \label{tab:nam_pcoc}
  \begin{tabular}{c c c c c }
    
    \hline\hline
    {Level} & {Model} & {PCOC} & {PCOC Diff} \\
    \hline
    \multirow{2}*{(0,1\%]}  & AEM & 1.1715  &  \\ 
                            & NAM & 1.1262  & -3.87\%\\ 
    
    \hline
    \multirow{2}*{(1\%,5\%]}  & AEM	& 0.9957  	 & \\	
    	                   & NAM	& 0.9863  & -0.94\%	 \\
    \hline
    \multirow{2}*{(5\%,20\%]}   & AEM	& 0.9076	 & \\	
                                & NAM	& 0.9019	& -0.63\%\\
    \hline
    \multirow{2}*{(20\%,100\%]} & AEM	& 0.8174	 & \\	
                                & NAM	& 0.8394	 & +2.69\%\\
    
    \hline
    \end{tabular}
    }
\end{table}

\subsection{Online A/B Test }
\begin{figure}[H]
    \centering
    \includegraphics[width=0.4\linewidth]{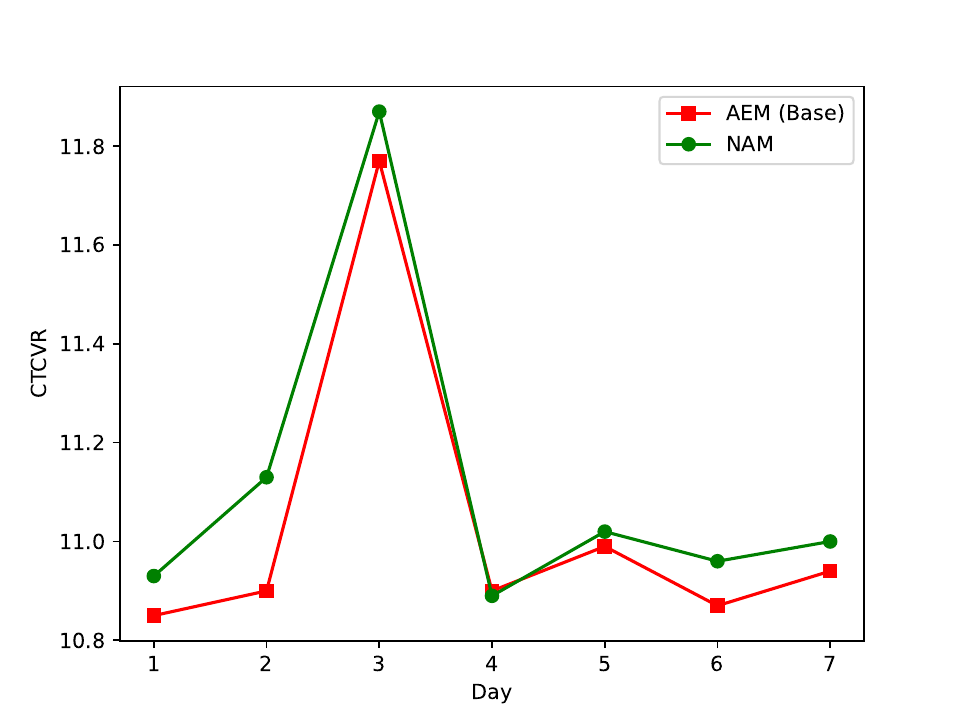}
    \caption{Online Performance of the proposed NAM}
    \label{fig:online_abtest_result}
\end{figure}

We conduct online A/B test in the product search system in Fliggy within 7 days of November 2024. Figure \ref{fig:online_abtest_result} illustrates the online performance of NAM compared to AEM (Base). The mean CTCVR has increased by 0.8\% (Given the substantial daily sales volume of Fliggy and hotel prices in the hundreds of yuan range, a 0.8\% increase in sales volume indicates nearly one million yuan increase in revenue.).

\section{Conclusion}
In this paper, we present three intuitions that are relevant to the personalized product search practices, and investigate these intuitions through data analysis. Based on these observations, we propose a universally applicable personalized and normalized attention model, which optimizes ”when to personalize” by employing an Inverse Item Frequency (IIF) based gating mechanism, as well as optimizes ”how to personalize” by
normalizing the attention mechanism from a global perspective. 
We further conduct comprehensive experimental analyses on an industrial dataset, where the results demonstrate the significant performance improvements of our proposed method over state-of-the-art baselines and multiple variants of attention mechanisms. Finally, this approach also achieved significant benefits in the online environment of product search in Fliggy.

% In this paper, based on three insightful findings, we proposed a universally applicable personalized and normalized attention model, which optimizes ”when to personalize” by employing an Inverse Item Frequency (IIF) based gating mechanism, as well as optimizes ”how to personalize” by normalizing the attention mechanism from a global perspective. We further conduct comprehensive experimental analyses on an industrial dataset, where the results demonstrate the significant performance improvements of our proposed method over state-of-the-art baselines and multiple variants of attention mechanisms.

\bibliographystyle{ACM-Reference-Format}
\balance
\bibliography{reference}

% \appendix
% \input{sections/appendix}

\end{document}